\documentclass[aps, prl, superscriptaddress, twocolumn, letterpaper, 10pt, amsfonts, amsmath, amssymb]{revtex4-2}

\usepackage{graphicx}
\usepackage{hyperref} 
\usepackage{xcolor}
\usepackage{soul}
\hypersetup{
    colorlinks,
    linkcolor={blue!80!black},
    citecolor={blue!80!black},
    urlcolor={blue!80!black}
}
\usepackage{physics}

\usepackage{placeins}

\usepackage{siunitx}
\usepackage{chemformula}

\usepackage{xcolor}
\usepackage{lipsum}

\definecolor{editcolor}{rgb}{1, 0.0, 0.0}

\begin{document}
\title{The Josephson diode effect in supercurrent interferometers}
\author{Rubén Seoane Souto}
\affiliation{Center for Quantum Devices, Niels Bohr Institute, University of Copenhagen, 2100 Copenhagen, Denmark}
\affiliation{Division of Solid State Physics and NanoLund, Lund University, S-22100 Lund, Sweden}
\author{Martin Leijnse}
\affiliation{Center for Quantum Devices, Niels Bohr Institute, University of Copenhagen, 2100 Copenhagen, Denmark}
\affiliation{Division of Solid State Physics and NanoLund, Lund University, S-22100 Lund, Sweden}
\author{Constantin Schrade}
\affiliation{Center for Quantum Devices, Niels Bohr Institute, University of Copenhagen, 2100 Copenhagen, Denmark}

\date{\today}

\begin{abstract}
A Josephson diode is a non-reciprocal circuit element that supports a larger dissipationless supercurrent in one direction than in the other. In this work, we propose and theoretically study a class of Josephson diodes based on supercurrent interferometers containing mesoscopic Josephson junctions, such as point contacts or quantum dots, which are not diodes themselves but possess non-sinusoidal current-phase relations. We show that such Josephson diodes have several important advantages, like being electrically tunable and requiring neither Zeeman splitting nor spin-orbit coupling, only time-reversal breaking by a magnetic flux. We also show that our diodes have a characteristic AC response, revealed by the Shapiro steps. Even the simplest realization of our Josephson diode paradigm that relies on only two junctions can achieve efficiencies of up to $\sim40\%$ and, interestingly, far greater efficiencies are achievable by concatenating multiple interferometer loops.
\end{abstract}

\maketitle
Non-reciprocal circuit elements underlie many modern applications of semiconductor electronics, such as current rectifiers, power converters and photodetectors. The simplest and most paradigmatic example of a non-reciprocal circuit element is the semiconductor diode that, conditioned on the polarity of an applied voltage bias, allows or prevents a \textit{dissipative current} flow. Driven by the need for energy-efficient circuitry, there has recently been mounting interest in extending the concept of a semiconducting diode to a `Josephson diode' (JD) \cite{ando2020observation,bauriedl2021supercurrent,daido2022intrinsic,shin2021magnetic,he2021phenomenological,Ilic_arXiv2022,pal2021josephson,Davydova_arXiv22,yuan2021supercurrent,Wu_arXiv2021,Diez_arXiv21,lin2021zero,scammell2021theory,Zhang_arXiv21,kononov2020one,chen2018asymmetric,Baumgartner_NatNano21,baumgartner2022effect,reynoso2012spin,yokoyama2013josephson,yokoyama2014anomalous,zazunov2009anomalous,brunetti2013anomalous,pal2019quantized,Halterman_arXiv21,strambini2021rectification,silaev2014diode} that, depending on the direction of a current bias, supports a larger  \textit{dissipationless supercurrent} in one direction than in the other. Promising applications of such a JD comprise superconducting (SC) electronic circuits with reduced power consumption and signal isolation in superconducting neural networks \cite{Schegolev_review2021,fuCMTJC}.

However, a major difficulty towards realizing a JD is that the current-phase relation (CPR) in time-reversal or inversion-symmetric Josephson junctions (JJs) is symmetric around zero SC phase difference, $I(\varphi)=-I(-\varphi)$. This property of the CPR ensures that the critical currents for the two current-bias directions, $I^{+}_{c}$ and $I^{-}_{c}$, are equal so that a JD effect is missing. To circumvent this symmetry constraint, previous works have observed and studied a diode effect in inversion-symmetry breaking superconducting thin films \cite{ando2020observation,bauriedl2021supercurrent,daido2022intrinsic,shin2021magnetic,he2021phenomenological,Ilic_arXiv2022}, JJs of finite-momentum pairing superconductors \cite{pal2021josephson,Davydova_arXiv22,yuan2021supercurrent}, JJs of multi-layered materials \cite{Wu_arXiv2021,Diez_arXiv21,lin2021zero,scammell2021theory,Zhang_arXiv21},  topological insulator JJs \cite{kononov2020one,chen2018asymmetric}, and JJs with a spin-orbit coupled two-dimensional electron gas subject to a Zeeman field \cite{Baumgartner_NatNano21,baumgartner2022effect,reynoso2012spin}. Moreover, it was also noted that a diode effect can arise in spin-orbit coupled nanowire JJs subject to a Zeeman field \cite{yokoyama2013josephson,yokoyama2014anomalous,zazunov2009anomalous,brunetti2013anomalous}, magnetic junctions \cite{pal2019quantized,Halterman_arXiv21,strambini2021rectification}, and domain wall SC channels \cite{silaev2014diode}.

 \begin{figure}[!b] \centering
\includegraphics[width=1\linewidth] {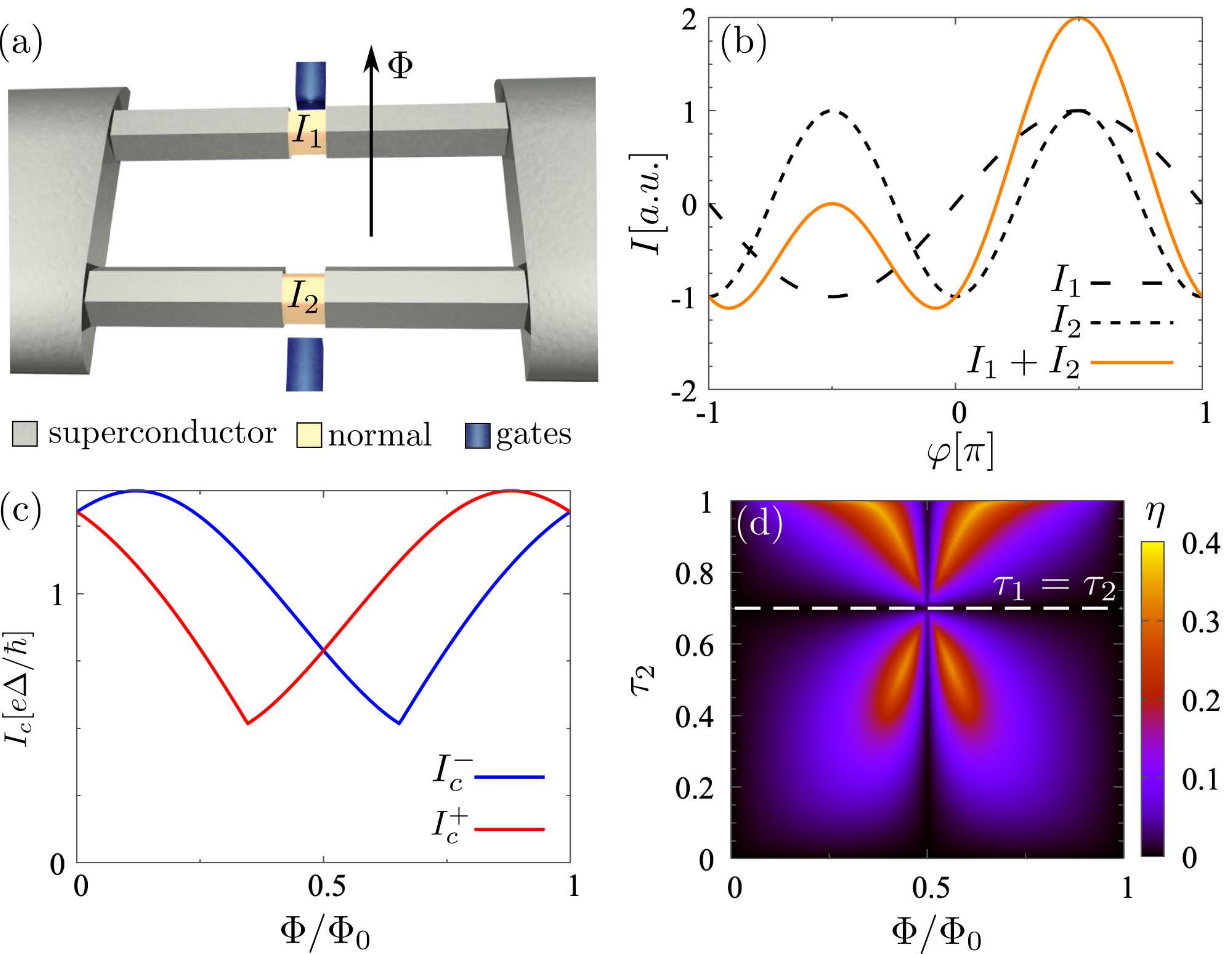}
\caption{(a) Minimal experimental setup for a JD interferometer comprising two JJs with non-sinusoidal CPRs. (b) Interference of different supercurrent harmonics, $I_{\ell=1,2}$, that gives rise to the JD effect. (c) Critical currents for two current-bias directions, $I^{\pm}_{c}$, in a QPC interferometer versus external flux, $\Phi$. (d) Diode efficiency, $\eta$, of a QPC interferometer versus external flux, $\Phi$, and transmission, $\tau_{2}$, of the second JJ. The transmission of the first JJ is $\tau_1=0.7$. 
}\label{Fig1}
\end{figure}

In this work, we consider an alternative platform for a JD in a supercurrent interferometer containing two or more mesoscopic JJs, such as point contacts or quantum dots. These JJs are not diodes themselves but have a high-harmonic content in their CPR. We show that such JDs have multiple advantages, such as their electrical tunability and requiring \textit{neither} a Zeeman splitting \textit{nor} finite spin-orbit coupling, only time-reversal breaking by a magnetic flux. For characterizing the JD effect, we use the diode efficiency, defined as
\begin{equation}
\eta=
\frac{
\Delta I_{c}
}
{I^{+}_{c}+I^{-}_{c}}
\quad\text{with}\quad
\Delta I_{c}=|I^{+}_{c}-I^{-}_{c}|.
\label{Eq1}
\end{equation}
While we show that this diode efficiency can reach $\sim40\%$ for the single-loop interferometer with two JJs, we also propose a route for systematically increasing the diode efficiency by concatenating multiple interferometers. Finally, we demonstrate that the JD effect also manifests itself in the AC response of our setup as asymmetric Shapiro steps.

\textit{Setup and minimal model.}
Our setup comprises a pair of mesocopic JJs ($\ell=1,2$) that connect two conventional SC leads to form a supercurrent interferometer, see Fig.\,\ref{Fig1}(a). To illustrate the origin of the JD effect in such an interferometer, we introduce a minimal model in which the CPRs of the two interferometer JJs include only a first harmonic and second harmonic contribution,
\begin{equation}
\begin{split}
I_{\ell}(\varphi)&=
I^{(1)}_{\ell}\sin(\varphi)+
I^{(2)}_{\ell}\sin(2\varphi).
\end{split}
\label{Eq2}
\end{equation}
Here, $\varphi$ is the SC phase difference and $I^{(m)}_{\ell}$ is the amplitude of the $m^{\text{th}}$ harmonic for the $\ell^{\text{th}}$ JJ. Upon piercing an external magnetic flux $\Phi$, through the area between the JJs, the total CPR of the interferometer reads,
\begin{equation}
I(\varphi)=
I_{1}(\varphi)+I_{2}(\varphi-2\pi\Phi/\Phi_0),
\label{Eq3}
\end{equation}
where $\Phi_{0}=h/2e$ denotes the magnetic flux quantum. The critical currents for the two current-bias directions are given by 
$I^{\pm}_{c}=\text{max}_{\varphi}[\pm I(\varphi)]$.

With the help of this minimal model, we can now explain how the JD effect arises in our proposed interferometer setup. For that purpose, we initially set $\Phi=\Phi_{0}/8$ and consider the case when one JJ comprises only a first harmonic, $I_{1}(\varphi)\propto\sin(\varphi)$, while the other JJ comprises only a second harmonic, $I_{2}(\varphi-2\pi\Phi/\Phi_0)\propto\cos(2\varphi)$. As shown in Fig.\,\ref{Fig1}(b), the two CPRs then interfere destructively for $-\pi<\varphi<0$ and constructively for $0<\varphi<\pi$. 
This interference effect, which behaves \textit{opposite} for positive and negative SC phase differences, gives rise to unequal critical currents for the two current bias directions, $I^{+}_{c}\neq I^{-}_{c}$, and thus to a JD effect.

This conceptual picture of the origin of the JD effect in our interferometer setup does not change qualitatively when adding additional harmonics to the JJs CPRs. In the previous situation when one of the JJs only comprised only a second harmonic so that $I^{(1)}_{2}=0$, the $I^{-}_{c}$ critical current corresponded to one of two minima within the range $-\pi<\varphi<0$, see Fig.\,\ref{Fig1}(b). When gradually turning on, for example, a finite first harmonic so that $I^{(1)}_{2}>0$, these two minima acquire different depth. However, the values of the global maximum and the global minimum of the CPR remain different. Hence, the JD effect persists also in this scenario.

{\it Quantum point contact interferometer.} 
So far, our minimal model helped us gain a conceptual picture of the emergence of the JD effect. A natural next step is to understand the experimental requirements for a JD. For that purpose, we consider more realistic setups. We first consider the simple case when the two nanowire JJs realize quantum point contacts (QPCs) with CPRs, $I_{\ell}(\varphi)$, that are given by \cite{beenakker1991universal,martinis2004superconducting},
\begin{equation}
\begin{split}
    I_{\ell}(\varphi)&=\frac{e\Delta^2\tau_{\ell}}{2\hbar}\frac{\sin(\varphi)}{\varepsilon_{\ell}(\varphi)}\,{\rm Tanh}\left[\varepsilon_{\ell}(\varphi)/2k_BT\right]\,,
    \\
    \varepsilon_{\ell}(\varphi)&=\Delta\sqrt{1-\tau_{\ell}\,\sin^2(\varphi/2)}\,.
   \label{Eq4}
\end{split}
\end{equation}
Here, $\varepsilon_{\ell}(\varphi)$ are the Andreev levels that emerge within the SC gap, $\Delta$, and that mediate the supercurrent. Moreover, $\tau_{\ell}$ are the transmission of the $\ell^{\text{th}}$ QPC JJ and $T$ is the temperature. Henceforth, we will assume $k_{B}T\ll\Delta$. In this situation, the CPR of the QPC JJ is approximately sinusoidal for small transmissions, $I_{\ell}(\varphi)\propto\sin(\varphi)$ if $\tau_{\ell}\ll1$. However, for large transmissions, $\tau_{\ell}\lesssim 1$, the CPR gets increasingly `skewed', thereby acquiring higher harmonic content, $I_{\ell}(\varphi)=
I^{(1)}_{\ell}\sin(\varphi)+
I^{(2)}_{\ell}\sin(2\varphi)+\dots$. In this regime, the QPC supercurrent interferometer realizes an approximate version of our minimal model.

Returning to our question on the experimental requirements for the JD, we proceed by computing the diode efficiency of the QPC interferometer as a function of the transmissions, $\tau_{\ell}$, and the external flux, $\Phi$. A representative result is depicted in Fig.\,\ref{Fig1}(d). It shows that the JD diode efficiency, $\eta$, comprises four lobes with a maximum efficiency of $\sim40\%$. Moreover, we find from the simulations that the JD requires only three simple requirements that need to be \textit{simultaneously} satisfied: 

(1) The external flux should not be an integer multiple of the half flux quantum, $\Phi\neq n\Phi_{0}/2$. If this condition is not satisfied, then 
$I(\varphi)=-I(-\varphi)$ and $I^{+}_{c}=I^{-}_{c}$, implying that the JD effect vanishes, see Fig.\,\ref{Fig1}(c).

(2) The junctions transmissions should not be equal, $\tau_{1}\neq\tau_{2}$. If this condition is not satisfied, then the CPR has the property $I(\varphi)=-I(-\varphi+2\pi\Phi/\Phi_0)$, implying $I^{+}_{c}=I^{-}_{c}$. The vanishing diode efficiency when $\tau_{1}=\tau_{2}$ is shown by dashed line in Fig.\,\ref{Fig1}(d).

(3) At least one of the JJ needs to be highly transmitting. This requirement ensures a sizable higher harmonic content in the total CPR.
If such a higher harmonic content is negligible (compared to the magnitude of the first harmonics), then the JD effect is again missing. 

Lastly, we note that while we focused on single-channel QPCs, the same requirements also apply to QPCs with multiple conduction channels. In that case, the requirement (2) generalizes to the condition that the amplitudes of the individual harmonics of the two CPRs from the interferometer JJs should not be pairwise equal.

\begin{figure}[!b] \centering
\includegraphics[width=1\linewidth] {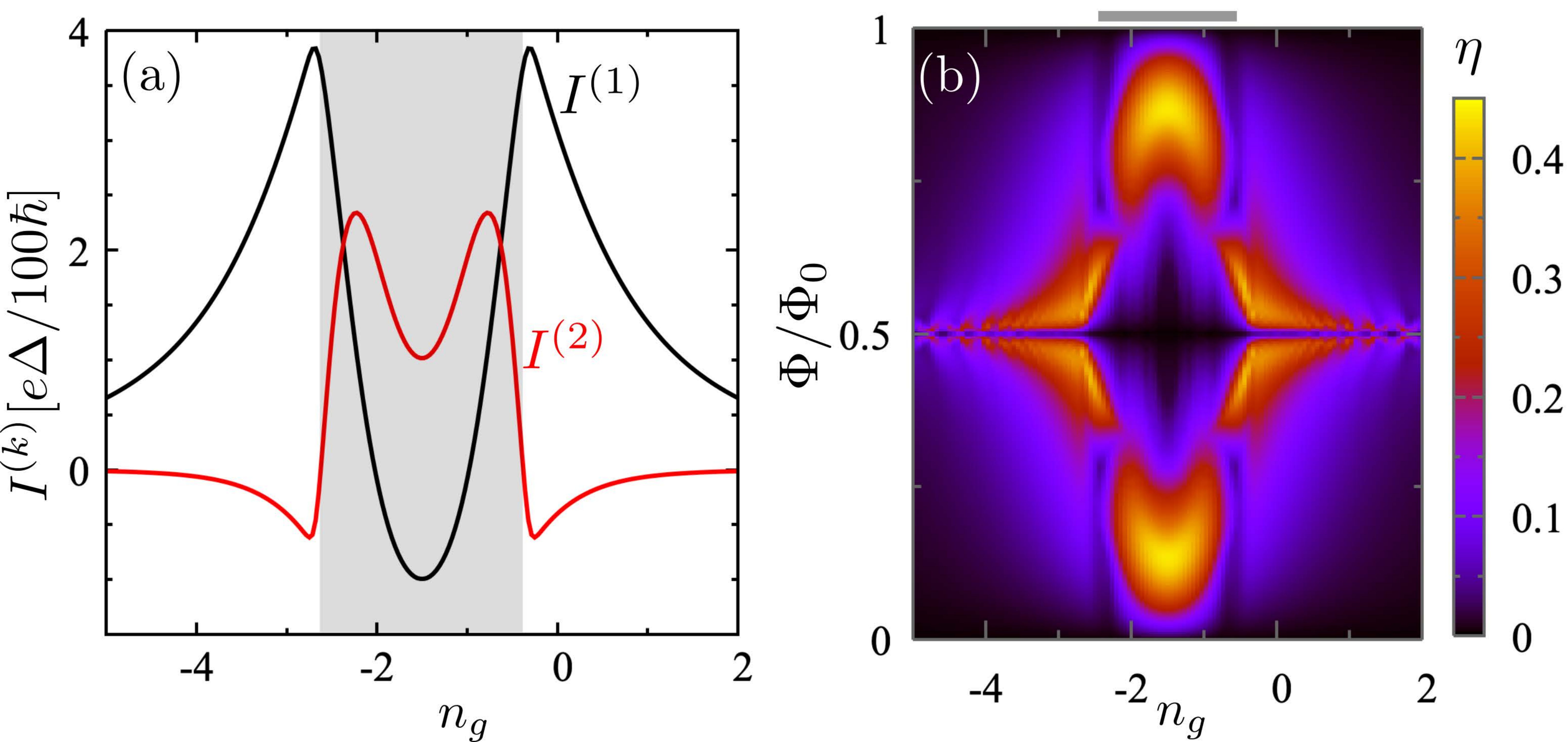}
\caption{(a) First and second harmonic, $I^{(1,2)}$, of a QD JJ CPR with $(U,t_{L},t_{R},T)=(3,2/3,2/3,0.001)\Delta$ versus offset charge, $n_{g}$. See \cite{Supplemental} for the detailed model. 
(b) Diode efficiency, $\eta$, for an interferometer with a QD JJ and QPC JJ versus 
flux, $\Phi$, and offset charge, $n_{g}$. Here, $\eta$ is optimized with respect to the QPC JJ transmission. QD parameters are as in (a). 
}\label{Fig2}
\end{figure}

\textit{Quantum dot interferometer.} In our discussion of the QPC interferometer, we have seen that a CPR with higher harmonic content is 
an important requirement for realizing the JD. Interestingly though, such higher harmonics can also emerge in nanowire JJs that are not highly-transmitting QPCs.  An example is a JJ realized by a spin-degenerate quantum dot (QD) with charging energy, $U$, that couples with tunneling amplitudes, $t_{L/R}$, to SC leads with SC gap, $\Delta$. Such a QD JJ can reverse the supercurrent when a gate voltage tunes the offset charge, $n_{g}$, on the dot \cite{spivak1991negative,van2006supercurrent}. Close to the supercurrent reversal, single Cooper pairs tunnel with approximately equal amplitude in the forward/reverse direction. As a result, the net current due to single Cooper pair tunneling, corresponding to the first harmonic, is reduced. In this regime, higher harmonics contributions become relevant. 

Motivated by this alternative mechanism to generate higher harmonics, we consider a second realization of the supercurrent interferometer with one low-transmitting QPC and one QD JJ. We first compute, assuming the zero-bandwidth approximation in the leads \cite{Allub_PRB1981,Estrada_PRL2018}, the CPR of the QD JJ. For the detailed model, see \cite{Supplemental}. Our results are shown in Fig.\,\ref{Fig2}(a). We find that the first harmonic is reduced close to the supercurrent reversal transitions, while the second harmonic exhibits an enhancement. If we now arrange the QD JJ and the low-transmitting QPC JJ in a supercurrent interferometer, the JD effect emerges again. We have computed the diode efficiency in Fig.\,\ref{Fig2}(b). As expected, the diode efficiency is maximized close to the supercurrent reversal transitions where the first and second harmonics have comparable magnitudes.

\begin{figure}[!b] \centering
\includegraphics[width=1\linewidth] {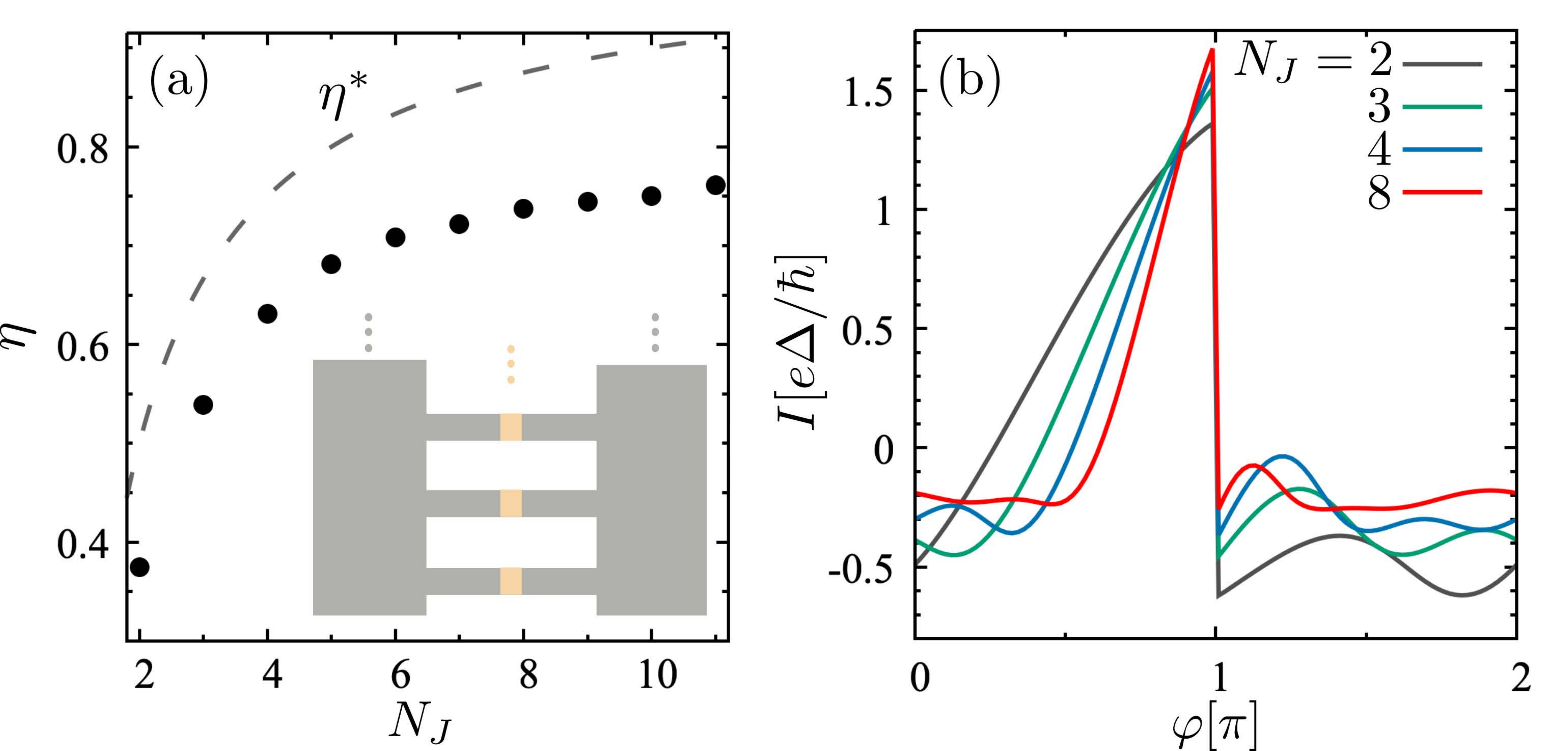}
\caption{(a) Maximum diode efficiency, $\eta$, for $N_{J}$ QPC JJ concatenated in an interferometer array (as shown in the inset). Dashed lines shows $\eta^\ast=(N_{J}-1)/N_{J}$. (b) CPR in the maximum diode efficiency configuration for various $N_{J}$.
}\label{Fig3}
\end{figure}

{\it Diode efficiency optimization.} 
Our previous JD implementations achieved efficiencies $\eta\sim 40\%$, see Fig.\,\ref{Fig1}(c) and Fig.\,\ref{Fig2}(b). We will now discuss how the diode efficiency can be systematically optimized. One conceptual approach is to include additional harmonics in the CPR that elevate the critical current in one direction and reduce it in the reverse one. As an example, we return to Fig.\,\ref{Fig1}(b) where $I(\varphi)\propto \sin(\varphi)-\cos(2\varphi)/\sqrt{2}$ and add a third harmonic $\propto -\sin(3\varphi)/\sqrt{3}$. This third harmonic will destructively/constructively interfere with the minima/maxima of the CPR, thereby enhancing the JD effect. Adding further harmonics so that $I(\varphi)\propto\sum^{N}_{n=1}\cos(n(\varphi-\pi/2))/\sqrt{n}$ leads to a CPR that approaches asymptotically $\eta=1$ for increasing $N$ \cite{Supplemental}.

To implement this conceptual approach experimentally, we propose an array of parallel QPC JJs arranged in an interferometer setup, see the inset of Fig.~\ref{Fig3}(a). We expect that tuning individually the interferometer fluxes and transmissions will create an optimized JD effect as discussed above. To find the maximum diode efficiencies, we use Monte Carlo maximization, randomly sampling the parameter space of transmissions and fluxes. We typically use $10^9$ sampling points, sufficient to find the maximum diode efficiency for $N_J\sim10$ QPCs. Results for the maximum efficiency are shown in Fig.\,\ref{Fig3}(a) as a function of $N_J$. We find that the efficiency grows monotonously with $N_J$, increasing faster for small $N_J$. These results suggest that \textit{already a small number} of concatenated loops is sufficient to significantly improve the diode efficiency.

However, we note that it is not possible to achieve perfect efficiency 
with a finite number of JJs, since there is no phase-independent term in a CPR. Hence, it is relevant to set bounds on the maximal efficiency achievable with $N_{J}$ JJs that are not diodes themselves. 
Two JJs achieve the highest diode efficiency when the supercurrent interference is optimal: the CPR maxima align, while the minimum of one junction aligns with the maximum of the other one. With this reasoning, the best upper bound we found is given by $50\%$ for two JJs. By concatenating $N_{J}$ JJs in the same way, the best upper bound we found for the diode efficiency is $\eta^\ast=(N_{J}-1)/N_{J}$ \cite{Supplemental}.

\begin{figure}[!b] \centering
\includegraphics[width=1\linewidth] {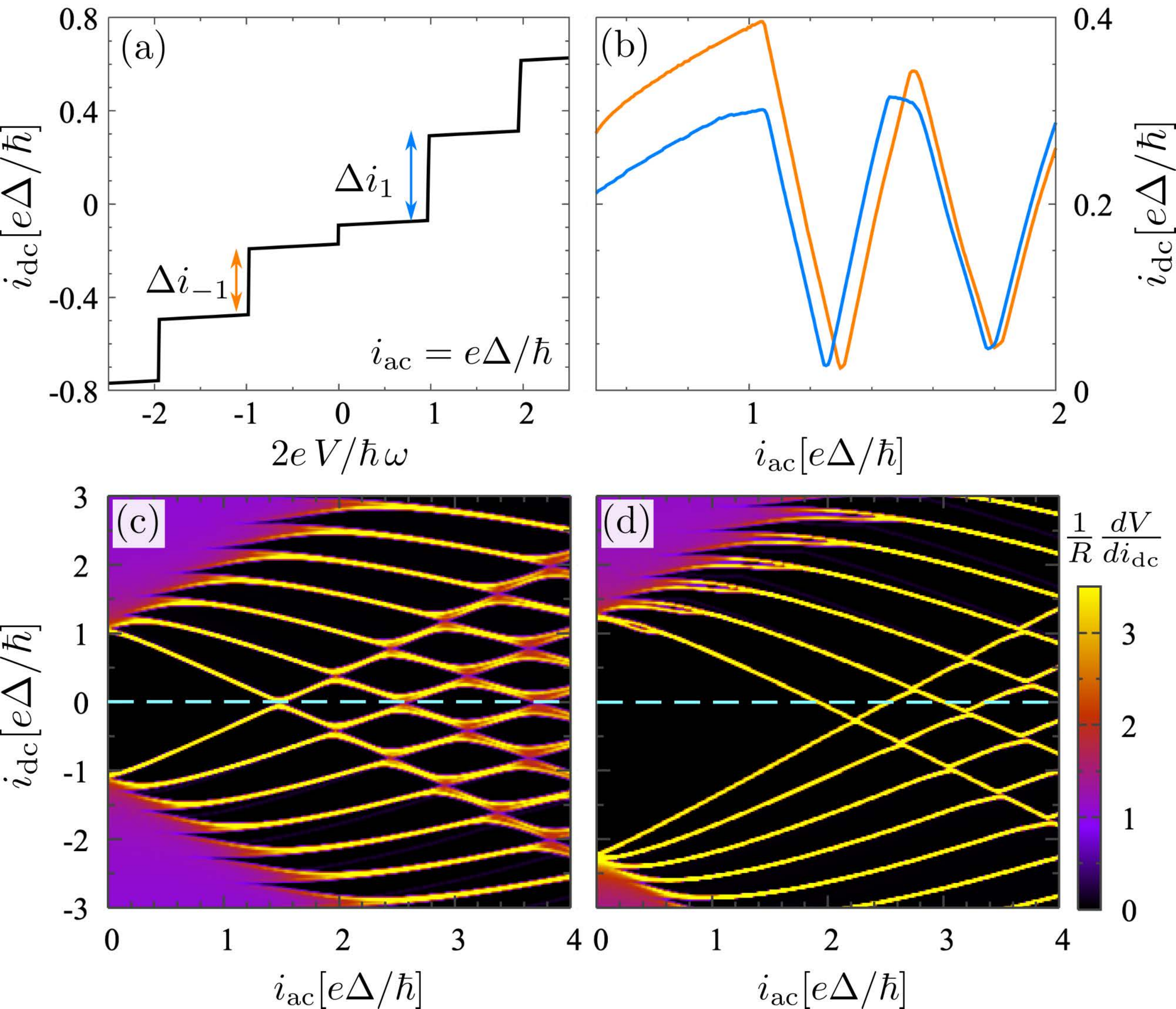}
\caption{(a) $i_{\text{dc}}-V$ curve of a AC-biased QPC interferometer, showing
Shapiro steps with step heights, $\Delta i_{n}$. (b) Step heights, $\Delta i_{\pm1}$, as a function of $I_{\text{ac}}$, showing a relative offset. (c) Differential resistance, $\text{d}V/\text{d}I_{\text{dc}}$, versus $I_{ac}$ and $I_{\text{dc}}$, for a QPC interferometer without JD effect, showing a symmetry around $I_{\text{dc}}=0$. (d) Same as (c) but for a QPC interferometer with JD effect. The symmetry around $I_{\text{dc}}=0$ is missing. Parameters used for all plots are $(\tau_{1},\tau_{2},\Phi/\Phi_{0},\omega)=(0.7,0.99,0.32,0.2\Lambda)$ with $\Lambda=(2eR/\hbar)(e\Delta/\hbar)$, except (c) uses $\tau_{2}=0.7$. 
}\label{Fig4}
\end{figure}

{\it Shapiro steps.} While we have so far explored the JD effect in DC-biased interferometers, it is interesting to ask if there are also manifestations of the JD effect in an interferometer with \textit{both} an applied DC- \textit{and} AC-bias. To address this question, let us consider for concreteness an interferometer with single-channel QPC JJs.
The time-dependence of the phase difference, $\varphi(t)$, and the voltage drop, $V(t)=(\hbar/2e)\dot{\varphi}(t)$, is then given by \cite{stewart1968current,mccumber1968effect,park2021characterization},
\begin{equation} 
\frac{\hbar}{2eR}\,\dot{\varphi}(t) + I(\varphi(t))= i_{\text{dc}} + i_{\text{ac}} \sin(\omega t).
\label{Eq5}
\end{equation}
Here, $i_{\rm dc}$ and $i_{\rm ac}$ are the amplitudes of the DC- and AC-bias. Moreover, $\omega$ is the driving frequency and $R$ is a shunt resistance in parallel to the interferometer.
For $I(\varphi)\propto \sin(\varphi)$, the primary consequence of Eq.\,\eqref{Eq5} is the appearance of voltage plateaus or `Shapiro steps' in the $i_{\rm dc}-V$ curve when $V=V_{n}\equiv 2n\hbar\omega/e$. Here, $n$ is an integer. We now want to understand how these Shapiro steps change for our JD interferometer. 

To identify the possible changes of the Shapiro pattern for a JD interferometer, we have solved Eq.\,\eqref{Eq5} numerically, leading to two notable findings:

(1) In the $i_{\rm dc}-V$ curve of a conventional JJ with sinusoidal CPR, the step size (or plateau length in $i_{\rm dc}$) of the $n^{\text{th}}$ Shapiro step at $V=V_{n}$ is equal to the step size of the $-n^{\text{th}}$ Shapiro step, $\Delta i_{n}=\Delta i_{-n}$. This equality follows from the property $i_{\text{dc}}(V)=-i_{\text{dc}}(-V)$. In a JD interferometer, this no longer holds. As shown in Fig.\,\ref{Fig4}(a), we find that the step sizes of the  $n^{\text{th}}$ and $-n^{\text{th}}$ Shapiro steps are, in general, unequal, $\Delta i_{n}\neq\Delta i_{-n}$. This unusual behavior of the Shapiro step size becomes particularly apparent when tuning $i_{\rm ac}$, which shows oscillations with a relative offset between $\Delta i_{n}$ and $\Delta i_{-n}$, see Fig.\,\ref{Fig4}(b). 

(2) For a conventional JJ with no JD effect, a plot of the differential resistance, $\text{d}V/\text{d}i_{\text{dc}}$, versus $i_{\rm ac}$ and $i_{\rm dc}$ shows a `fan-like' pattern with sharp peaks marking the Shapiro steps. As shown in Fig.\,\ref{Fig4}(c), in the absence of the JD effect, the peaks in the differential resistance are mirror-symmetric with respect to $i_{\rm dc}=0$ and the separation between neighboring peaks on a fan (for fixed $i_{ac}$) gives the Shapiro step size. In a JD interferometer, we find that the differential resistance peaks lack the symmetry around $i_{\rm dc}=0$, see Fig.\,\ref{Fig4}(d). This unusual behavior arises due to the unequal critical currents for the two current-bias directions, so that the peaks emerge at different $i_{\rm dc}$ values for $i_{\rm ac}=0$. Moreover, we also note that the separation between neighboring peaks is different for positive/negative DC current bias directions, which reconfirms our previous finding that $\Delta i_{n}\neq\Delta i_{-n}$.

\begin{figure}[!b] \centering
\includegraphics[width=1\linewidth] {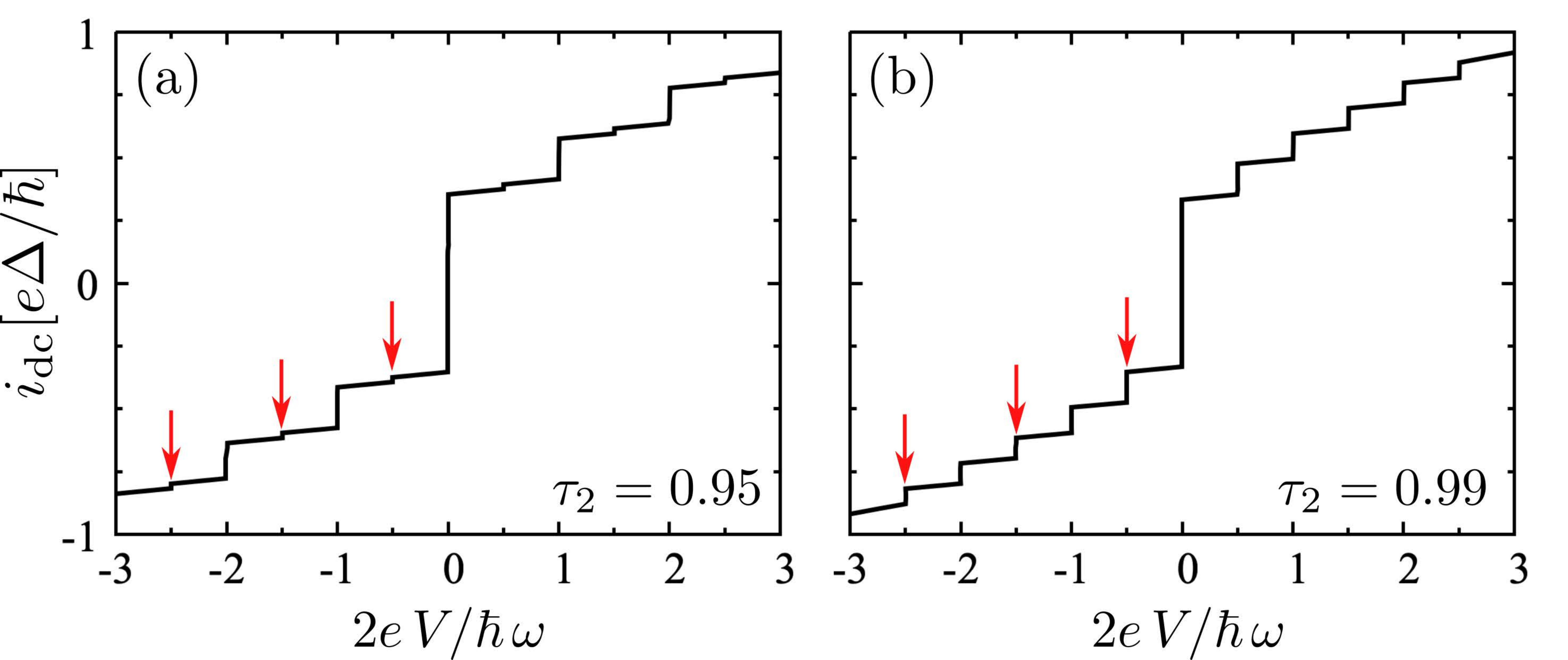}
\caption{Shapiro steps in the $i_{\rm dc}$-V curve for an DC- and AC-biased QPC interferometer 
with drive amplitude, $i_{\rm ac}=0.4e\Delta/\hbar$, and $\Phi/\Phi_{0}=1/2$ (a) For $(\tau_{1},\tau_{2})=(0.99,0.95)$, fractional Shapiro steps appear, highlighted by red arrows, which signal the importance of higher harmonics. (b) For $(\tau_{1},\tau_{2})=(0.99,0.99)$, the fractional Shapiro steps have the same step-heights as the integer Shapiro steps. 
}\label{Fig5}
\end{figure}

Finally, we remark that earlier works have found that `fractional Shapiro steps' can arise in QPC JJs at voltages $V=V_{n/m}\equiv 2(n/m)\hbar\omega/e$ due to the higher harmonic content in the CPR \cite{cuevas2002subharmonic,duprat2005phase,chauvin2006superconducting}. Interestingly, such fractional Shapiro steps can emerge in our setup in an even more pronounced way if $\tau_{\ell}\lesssim1$. 
For example, by setting $\Phi=\Phi_{0}/2$ and tuning the JJs from an unbalanced ($\tau_{1}\neq\tau_{2}$) to a balanced situation ($\tau_{1}=\tau_{2}$), we find an \textit{exact} doubling of the Shapiro steps, with
all integer steps having the same height as half-integer steps, see Fig.\,\ref{Fig5}. This exact doubling of Shapiro steps is due to pure double-Cooper pair tunneling, leading to the formation of an effective $\cos(2\varphi)$ Josephson potential. In future experiments, the transition to a doubled Shapiro step pattern could be useful for calibrating parity-protected qubits \cite{larsen2020parity,gyenis2021moving,schrade2021protected,maiani2022entangling}, which require a  $\cos(2\varphi)$ Josephson potential to robustly store quantum information. 

{\it Conclusions.} In this work, we considered an alternative paradigm for realizing JDs based on supercurrent interferometers containing JJs with a non-sinusoidal CPR. 
We showed that such JD interferometers can achieve sizable diode efficiencies that can be optimized by concatenating interferometer loops. Finally, we also discussed a manifestation of the JD effect in the Shapiro pattern, which could 
open the way to a new series of diode experiments based on the AC Josephson effect.

We acknowledge helpful discussions with J. M. Chavez-Garcia and M. Kjaergaard. 
We acknowledge funding from the Swedish Research Council (VR) and the European Research Council (ERC) under the European Union’s Horizon 2020 research and innovation programme under Grant Agreement No. 856526, and Nanolund. We also gratefully acknowledge funding from the Microsoft Corporation.

\newpage
\begin{widetext}

\begin{center}
\large{\bf Supplemental Material to `The Josephson diode effect in supercurrent interferometers' \\}
\end{center}
\begin{center}
Rubén Seoane Souto$^{1,2}$, Martin Leijnse$^{1,2}$, and Constantin Schrade$^{1}$
\\
{\it $^{1}$Center for Quantum Devices, Niels Bohr Institute, University of Copenhagen, 2100 Copenhagen, Denmark}
\\
{\it $^{2}$Division of Solid State Physics and NanoLund, Lund University, S-22100 Lund, Sweden}
\end{center}
In the Supplemental Material, we provide details on the model for the quantum dot interferometer, the diode efficiency optimization, and the fractional Shapiro steps. 
\end{widetext}

\setcounter{figure}{0}
\setcounter{equation}{0}
\twocolumngrid
\renewcommand\thefigure{S\arabic{figure}}

\section{Quantum dot interferometer}
In this section, we provide more details on the model for the quantum dot (QD) interferometer, which we discussed in the main text.
\subsection{Quantum dot Hamiltonian}
First, we introduce the model of an interacting QD tunnel-coupled to superconducting (SC) leads 
described by the Hamiltonian,
\begin{equation}
    H=H_{QD}+H_{l}+H_T\,.
    \label{Eq:QDmodel}
\end{equation}

The first contribution to the Hamiltonian, $H$, describes the spin-degenerate, interacting QD and is given by,
\begin{equation}
    H_{QD}=n_g\sum_{\sigma}n_{\sigma}+Un_\uparrow n_\downarrow\,,
    \label{H_QD}
\end{equation}
where $n_\sigma=d^{\dagger}_{\sigma}d_{\sigma}$ is the number operator on the QD with $d_{\sigma}$ annihilating a QD electron with spin $\sigma=\uparrow,\downarrow$. Moreover, $n_g$ denotes the offset charge and $U$ is the on-site Coulomb interaction strength.

The second contribution to the Hamiltonian, $H$ in Eq. \eqref{Eq:QDmodel}, describes the SC leads and reads,
\begin{equation}
    H_{l}=\sum_{k\nu\sigma}\xi_{k\nu\sigma}c^{\dagger}_{k\nu\sigma}c_{k\nu\sigma}+\sum_{k\nu}\Delta_{k\nu}c^{\dagger}_{k\nu\uparrow}c^{\dagger}_{k\nu\downarrow}+{\rm H.c.}\,,
    \label{H_leads}
\end{equation}
where $c_{k\nu\sigma}$ is the annihilation operator for an electron in lead $\nu=L,R$ with spin $\sigma$ and quasi-momentum $k$. In the normal state of the leads, the dispersion of such an electron is given by $\xi_{k\nu\sigma}$. In the SC state of the leads, $\Delta_{k\nu}$ is magnitude of the SC order parameter.

Finally, the tunneling Hamiltonian that connects the QD to the SC leads is given by, 
\begin{equation}
    H_{T}^{}=\sum_{\nu\sigma}\left|t_{k\nu}\right|e^{i\varphi_\nu}\Delta_\nu c^{\dagger}_{k\nu\sigma}d_{\sigma}+{\rm H.c.}\,,
    \label{H_leads}
\end{equation}
where $t_{k\nu}$ is denotes the tunneling amplitude and $\varphi_\nu$ is the phase of the SC order parameter on SC lead $\nu$. We define the phase difference between the SC leads as $\varphi=\varphi_L-\varphi_R$.

\begin{figure}[b!]\centering
\includegraphics[width=1\linewidth] {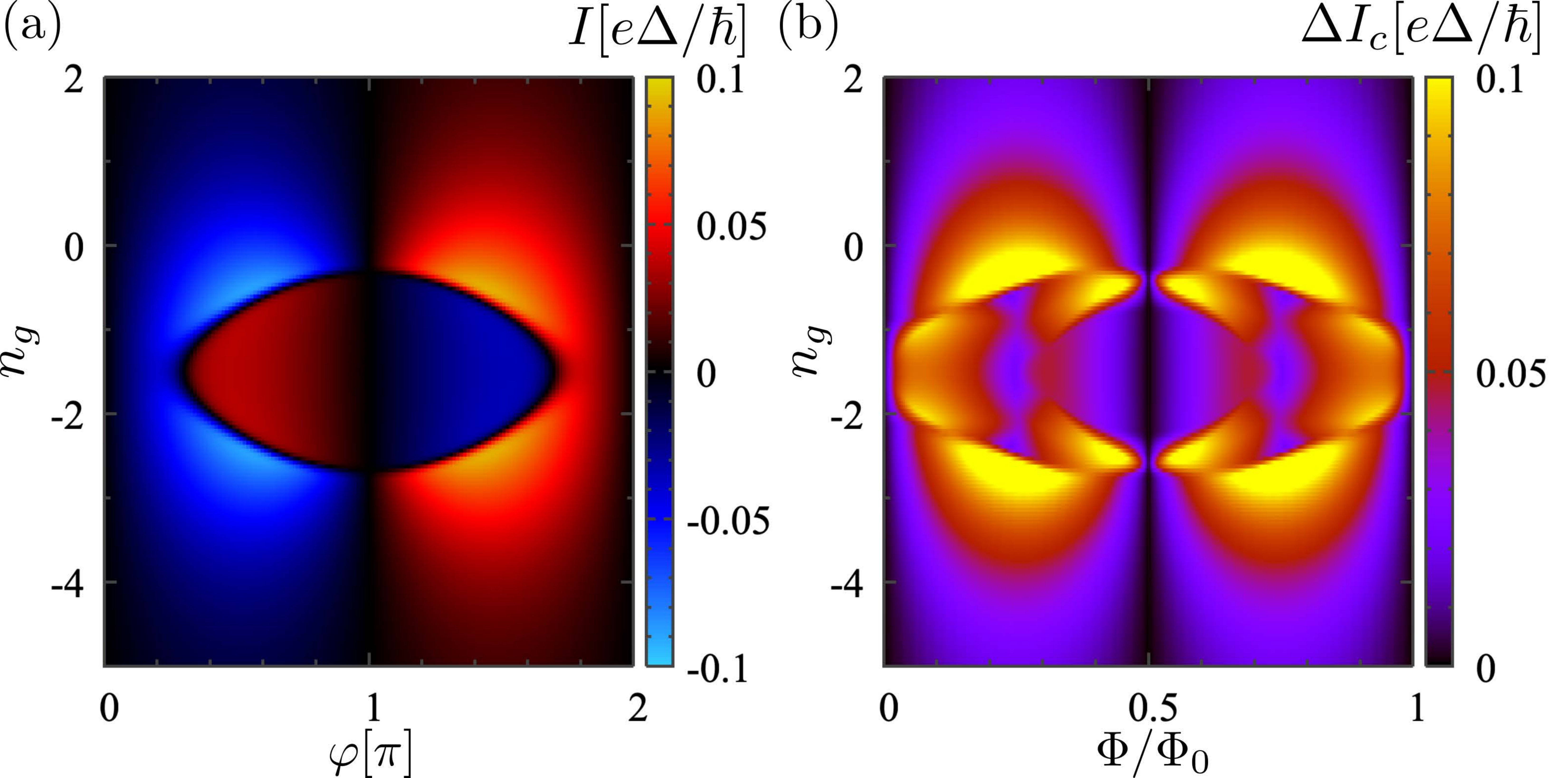}
\caption{(a) Supercurrent, $I$, in a QD JJ as a function of the SC phase difference, $\varphi$, and the offset charge, $n_{g}$. (b) Critical current asymmetry, $\Delta I_{c}$, as a function of the external flux, $\Phi$, and the offset charge, $n_{g}$. For both panels the parameters are $(U,t_{L},t_{R},T)=(3,2/3,2/3,1/100)\Delta$. 
}\label{S4}
\end{figure}

\subsection{Zero-bandwidth approximation}
There are different approaches for computing the supercurrent in a QD Josephson junction (JJ) such as perturbation theory in the tunneling amplitudes \cite{SpivakSM,Schrade2017SM,Rancic2019SM} or numerical renormalization group methods \cite{ChoiSM,Karrasch2008SM}. Here, we opt for an approach called the `zero-bandwith model'. This approximation, which has been shown to produce a good qualitative understanding of JJs with interacting QDs \cite{AllubSM,SaldanaSM}, approximates the SC leads by a single quasiparticle level. The Hamiltonian of the SC leads and the tunneling Hamiltonian thus take on the simplified form,
\begin{equation}
    \begin{split}
    H_{l}^{ZBM}&=\sum_{\nu}\Delta_\nu c^{\dagger}_{\nu\uparrow}c^{\dagger}_{\nu\downarrow}\,,
    \\
    H_{T}^{ZBM}&=\sum_{\nu\sigma}\left|t_\nu\right|e^{i\varphi_\nu}\Delta_\nu c^{\dagger}_{\nu\sigma}d_{\sigma}+{\rm H.c.}.
    \end{split}
\end{equation}
Here, zero energy corresponds to the center of the SC energy gap.

\begin{figure}[b!]\centering
\includegraphics[width=1\linewidth] {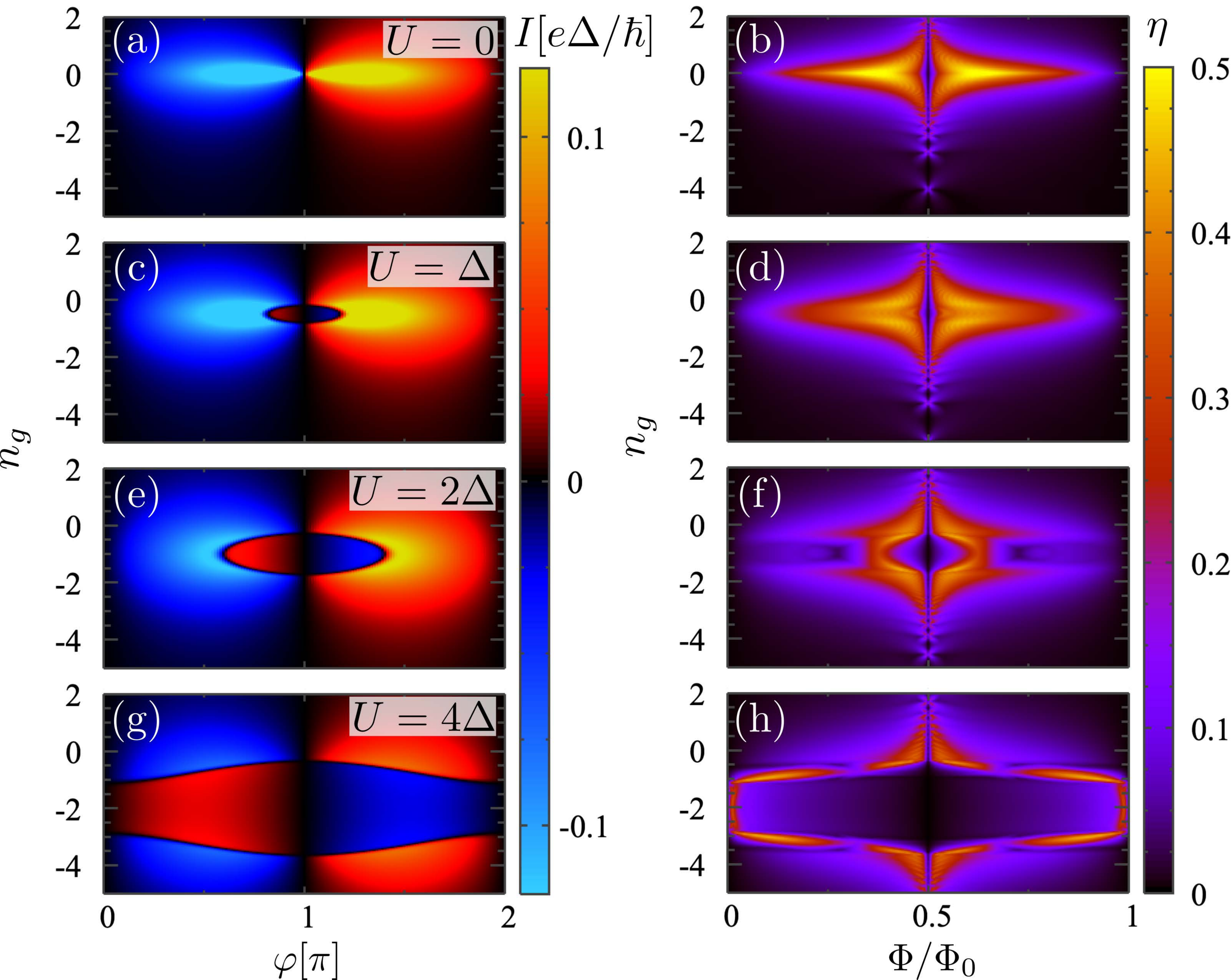}
\caption{Left panels: Supercurrent, $I$, as a function of the
SC phase difference, $\varphi$, and the offset charge, $n_g$, for 
various values of the interaction strength $U$. Right panels: Diode efficiency, $\eta$, as a function of the external flux, $\Phi$, and the offset charge, $n_g$ for the same interaction strength values, $U$, as in the left panels. System parameters are $(t_{L},t_{R},T)=(2/3,2/3,1/100)\Delta$.
}\label{S5}
\end{figure}

\subsection{Current-phase relation}
To compute the current-phase relation (CPR) of the QD JJ, we first perform a numerical exact diagonalization of the Hamiltonian, $H^{ZBM}$, given by, 
\begin{equation}
    H^{ZBM}=H_{QD}+H^{ZBM}_{l}+H^{ZBM}_T\,.
\end{equation}
From the energy levels of this Hamiltonian, denoted by $E_{j}(\varphi)$, we can compute the free energy at temperature $T$ as
\begin{equation}
    F(\varphi)=-k_B T\log\left(\sum_j e^{-E_j(\varphi)/k_BT}\right)\,.
\end{equation}
From this expression of the free energy, we obtain the CPR of the QD JJ as,
\begin{equation}
    I(\varphi)=\frac{2e}{\hbar}\frac{\partial F(\varphi)}{\partial \varphi}\,.\
    \label{current_F}
\end{equation}
The CPR for the QD JJ is shown in Fig. \ref{S4}(a) for the same parameters shown in Fig.\,3 of the main text. We see that there are two supercurrent reversal transitions as a function of the offset charge, $n_g$. Close to these transitions the CPR shows several sign changes within a period, resulting in a sizable higher harmonic content.

\subsection{Quantum dot interferometer}
In the main text, we considered a QD junction in parallel to a QPC junction. The diode efficiency, $\eta$, as a function of the offset charge, $n_{g}$, and the external flux, $\Phi$, 
reached a maximum of $\sim40\%$. 

\subsubsection{Critical current asymmetry}

We now want to complement the results of the main text with a numerical calculation of the critical current asymmetry, $\Delta I_{c}$, as a function of the the offset charge, $n_{g}$, and the external flux, $\Phi$. Our results are shown in Fig.\,\ref{S4}(b). We see that the critical current asymmetry is maximal in the vicinity of the supercurrent reversal transition. In this region, the CPR depicted in Fig.\,\ref{S4}(a) shows a strong non-sinusoidal behavior, characterized by a strong higher harmonic contribution, see Fig.\,2(a) in the main text.

\begin{figure}[b!] \centering
\includegraphics[width=1\linewidth] {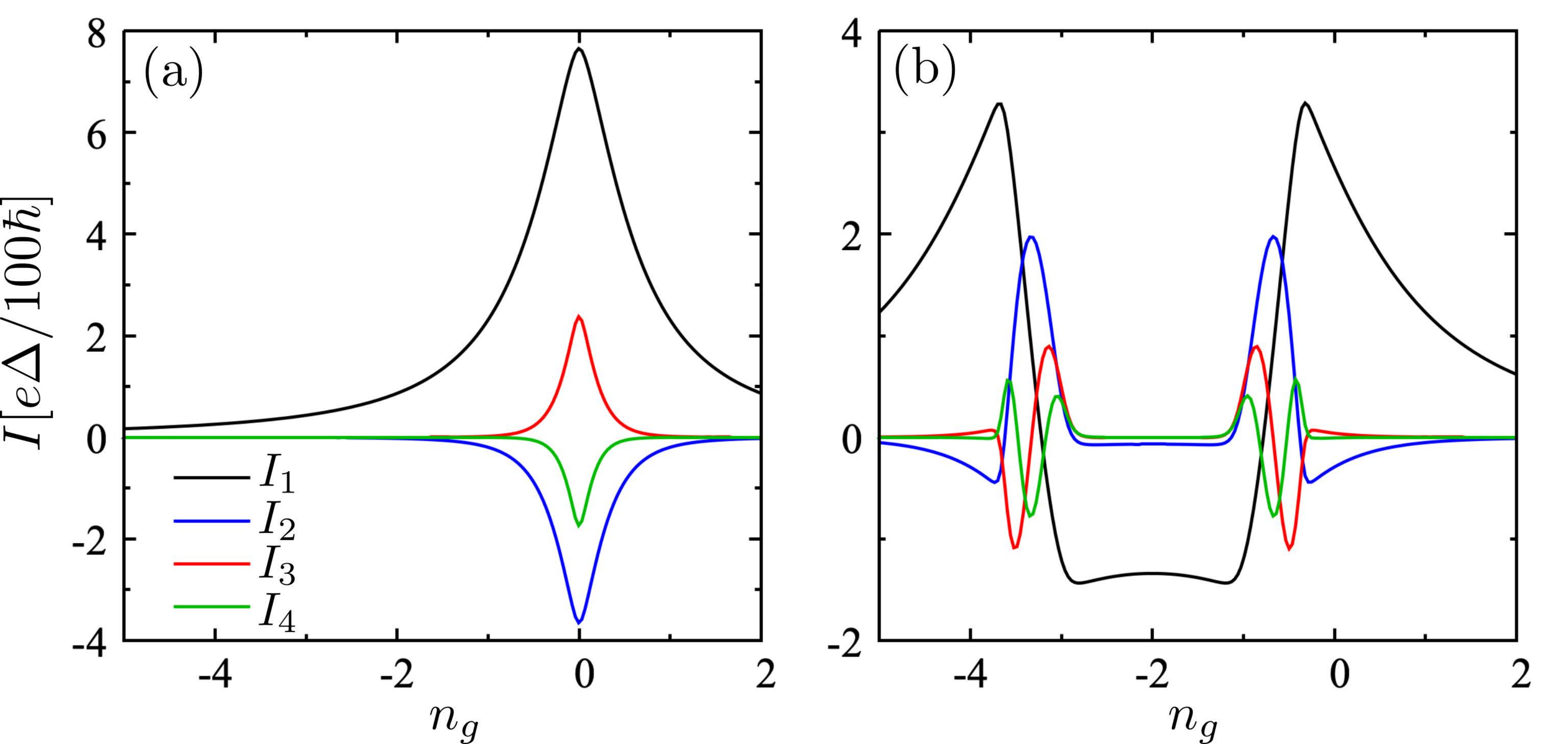}
\caption{(a) $n^{\text{th}}$ harmonic (for $n=1,\dots,4$) of a QD JJ as a function of the offset charge, $n_g$, when the Coulomb interaction strength on the QD vanishes, $U=0$. (b) Same as in (a) but for a finite Coulomb interaction strenght, $U=4\Delta$. The remaining parameters for both plots are $(t_{L},t_{R},T)=(2/3,2/3,1/100)\Delta$.
}\label{S6}
\end{figure}

\subsubsection{Effects of the quantum dot interaction strength}

We also want to provide further details on the on-site Coulomb interaction strength, $U$, modifies the JD effect. Our results are shown in Fig.\,\ref{S5} and Fig.\,\ref{S6}.

For the non-interacting case when $U=0$, 
we find that a sizable JD effect arises when the QD JJ is highly transmitting close to $n_{g}=0$, see Fig.\,\ref{S5}\,(a) and (b). In this regime, the CPR of the QD JJ exhibits sizable contributions from higher harmonics, as shown in Fig. \ref{S6}\,(a).

For the interacting case when $U>0$, we find that the effect of the interactions is two-fold: (1) The electron-hole symmetric point, defined by $I(\varphi,n_{g}-n^{*}_{g})=I(\varphi,-(n_{g}-n^{*}_{g}))$, shifts from $n^{*}_{g}=0$ to $n^{*}_{g}=-U/2$. (2) In the vicinity of the electron-hole symmetric point a supercurrent reversal transition emerges, see Fig.\,\ref{S6}\,(c),(e),(g). This transition amplifies the importance of higher harmonics in the CPR, as shown in Fig.\,\ref{S6}\,(a), and yields a JD effect with sizable efficiency, as shown in Fig.\,\ref{S6}\,(d),(f),(h).

\section{Diode efficiency optimization}
\label{Sec::IdealDiode}

In this section, we provide more details on the diode efficiency optimization
discussed in the main text. 

\subsection{Conceptual idea for improving the diode efficiency}

\begin{figure} \centering
\includegraphics[width=1\linewidth] {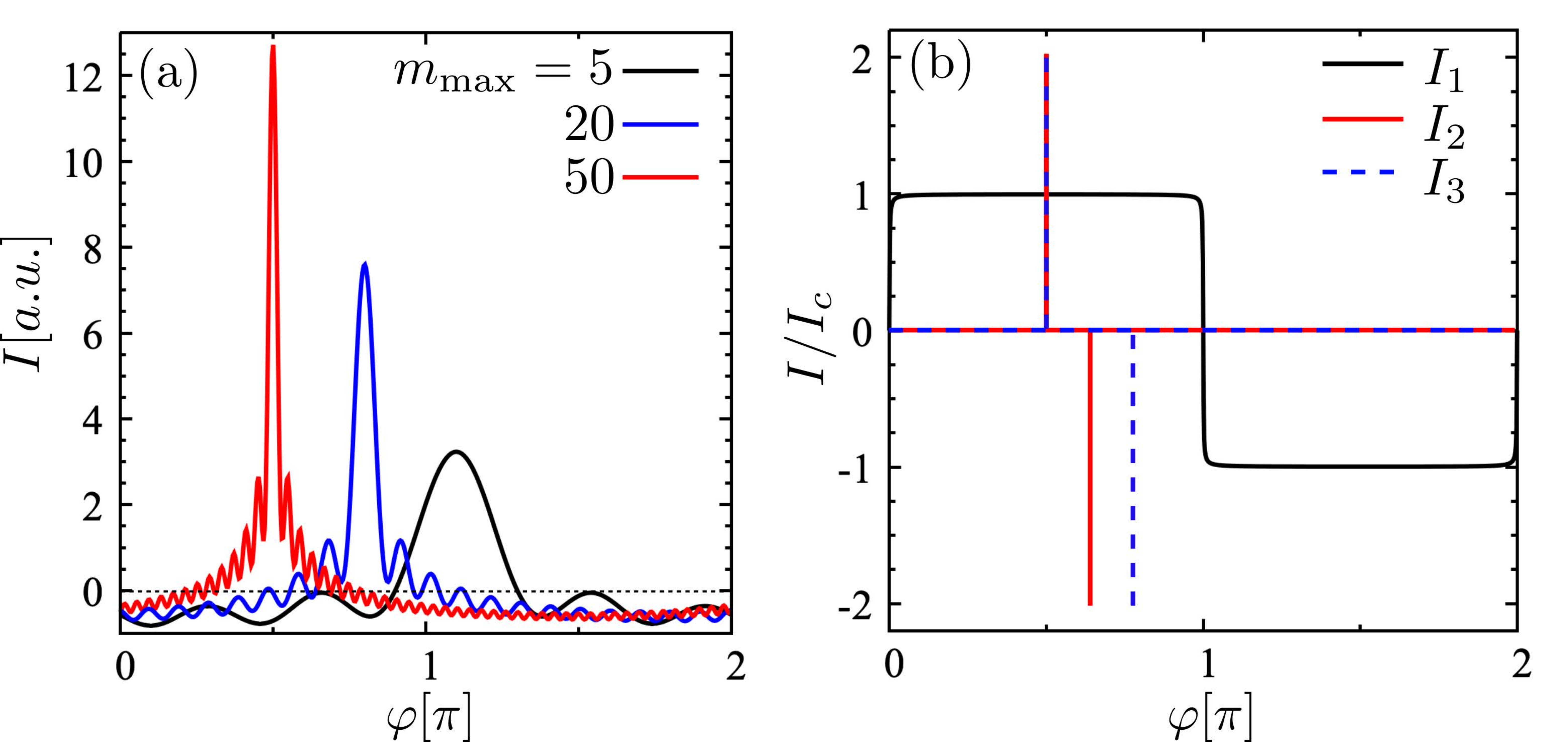}
\caption{(a) Plot of the CPR given by Eq.\,\eqref{ideal_diode} that asymptotically approaches $\eta=1$. (b) Example CPRs used for the construction of the
bound, $\eta^{*}$. 
}\label{S2}
\end{figure}

In the main text, we argued that it is not possible to realize a $\eta=1$ SC diode with a finite number of JJs because the CPR of JJ lacks a phase-independent (`$0^{\text{th}}$ harmonic') term. We have, therefore, proposed 
a conceptual example of a CPR that asymptotically approaches $\eta=1$, 
\begin{equation}
    I(\varphi)=\sum^{m_{\text{max}}}_{m=1}\frac{{\rm cos}[m(\varphi-\pi/2)]}{\sqrt{m}}\,.
    \label{ideal_diode}
\end{equation}
To further clarify the behavior of the CPR, we represent this CPR for 
values of $m_{\text{max}}$ in Fig. \ref{S2}\,(a). We see that upon increasing $m_{\text{max}}$, the CPR develops a sharp peak at $\varphi=\pi/2$, since
\begin{equation}
    I(\varphi=\pi/2)=\sum^{m_{\text{max}}}_{m=1}\frac{1}{\sqrt{m}}\,
\end{equation}
diverges upon increasing $m_{\text{max}}$. At the same time, the CPR approaches zero upon increasing $m_{\text{max}}$ at $\varphi=3\pi/2$ because 
\begin{equation}
    I(\varphi=3\pi/2)=\sum^{m_{\text{max}}}_{m=1}\frac{(-1)^{m}}{\sqrt{m}}\,
\end{equation}
converges to zero upon increasing $m_{\text{max}}$. 

\subsection{Concatenation of interference loops}
In the main text, we showed that the concatenation of multiple interference loops 
can significantly improve the diode efficiency. In Fig.\,3(b) of the main text, we have plotted the CPRs that yield the maximum diode efficiency for an array with $N_{J}$ JJs. 
The CPR of such an array is of the form,
\begin{equation}
    I(\varphi)=\sum_{\ell=1}^{N_J}I_{\ell}(\varphi-2\pi(\Phi^{(\ell)}-\Phi^{(\ell-1)})/\Phi_0)\,,
\end{equation}
where $\Phi^{(\ell)}$ is the flux piercing through the $\ell^{\text{th}}$ interferometer loop and $\Phi^{(0)}=0$. In Table \ref{table1} we provide the parameters the best JD (maximal $\eta$) found, where we fixed the transmission and the phase of one of the junctions to improve the convergence of our Monte Carlo maximization code. 

\begin{widetext}

\begin{table}[!t]
\centering
\caption{Parameters, $\{\tau_{\ell},2\pi(\Phi^{(\ell)}-\Phi^{(\ell-1)})/\Phi_0\}$, used to generate the CPRs in Fig.\,3(b) of the main text for $N_{J}$ JJs that are concatenated in an interferometr array. Here, $\Phi^{(\ell)}$ is the external flux that pierces through the $\ell^{\text{th}}$ interferometer loop.}
\begin{tabular}[t]{|l|c|c|c|c|c|c|c|c|c|c|c|c|c|c|cccc}
\hline
$N_J$ & 2 & 3 & 4 & 5 & 6 & 7 & 8 & 9 & 10 & 11\\
\hline
1 & $1.0,0.0$ & $1.0,0.0$& $1.0,0.0$& $1.0,0.0$ & $1.0,0.0$ & $1.0,0.0$ & $1.0,0.0$ & $1.0,0.0$ & $1.0,0.0$ & $1.0,0.0$\\
2 & $0.746,4.22$ & $0.833,4.90$& $0.9,5.21$& $0.891,0.55$& $0.935,4.50$ & $0.953,5.43$ & $0.894,0.48$ & $0.935,5.75$ & $0.951,5.64$ & $0.947,5.74$\\
3 & - & $0.828,3.02$& $0.888,2.37$& $0.887,1.83$& $0.914,0.66$ & $0.943,1.55$ & $0.884,4.53$ & $0.924,1.54$ & $0.893,1.50$ & $0.935,1.46$\\
4 & - & -& $0.823,3.78$& $0.879,4.14$& $0.865,3.35$ & $0.888,2.56$ & $0.780,3.82$ & $0.845,2.49$ & $0.815,2.37$ & $0.835,4.85$\\
5 & -  & - & - &  $0.877,3.13$ & $0.814,1.81$ & $0.687,4.62$ & $0.697,2.60$ & $0.811,4.68$ & $0.799,4.58$ & $0.797,2.38$\\
6 & -  & - & - & - & $0.635,2.43$ & $0.681,3.91$ & $0.677,1.30$ & $0.700,3.35$ & $0.724,3.43$ & $0.748,4.05$\\
7 & -  & - & - & - & - & $0.652,3.62$ & $0.651,1.76$ & $0.543,3.26$ & $0.544,4.51$ & $0.615,3.02$\\
8 & -  & - & - & - & - & - & $0.592,3.10$ & $0.541,4.54$ & $0.336,3.00$ & $0.507,3.08$\\
9 & -  & - & - & - & - & - & - & $0.106,4.45$ & $0.255,3.33$ & $0.481,5.02$\\
10 & -  & - & - & - & - & - & - & - & $0.097,2.86$ & $0.356,2.73$\\
11 & -  & - & - & - & - & - & - & - & - & $0.143,2.74$\\
\hline
$\eta$ & 0.374  & 0.538 & 0.631 & 0.681 & 0.708 & 0.722 & 0.737 & 0.744 & 0.750 & 0.761\\
\hline
\end{tabular}
\label{table1}
\end{table}%
\end{widetext}

\subsection{Diode efficiency bound}
In the main text, we argued that $\eta^{*}=(N_{J}-1)/N_{J}$ can provide an upper bound to the diode efficiency of $N_{J}$ JJs that are not diodes themselves. Here, we present more details on how we arrived at this bound with the help of a geometrical argument: 

Given the $N_{J}$ JJs, we assume that the the first Josephson junction (JJ) hosts a flat CPR with $I_{c}^{+}=I_{c}^{-}\equiv I_c$, black line in Fig. \ref{S2} (b). To achieve a high efficiency, the supercurrent maximum and minimum of a second JJ should align with maximum of the first one,  see Fig. \ref{S2} (b). The maximal efficiency is achieved for a critical current of the second JJ of $2I_c$. The sum of the described CPRs leads to $I_{c}^{+}=3I_{c}^{-}=3I_c$, leading to $\eta=0.5$ when $N_{J}=2$. It is possible to continue adding more peaked Josephson elements to increase the efficiency. The requirement is to align the peaks in one direction, while they should not overlap in the other one, dashed blue line in Fig. Fig. \ref{S2}. The critical currents in both directions scales with the number of JJs as
\begin{eqnarray}
    I_{c}^{+}=(2N_J-1)I_c\,,\qquad I_{c}^{-}=I_c\,,
\end{eqnarray}
leading to an upper bound for the diode efficiency
\begin{eqnarray}
    \eta^{*}=\frac{N_J-1}{N_J}\,.
\end{eqnarray}
We note that we have checked that $\eta^{*}$ also provides a bound for the CPR 
in Eq. \eqref{ideal_diode} when summing up a finite number of harmonics $m_{\rm max}$.

\begin{figure}[!t] \centering
\includegraphics[width=1\linewidth] {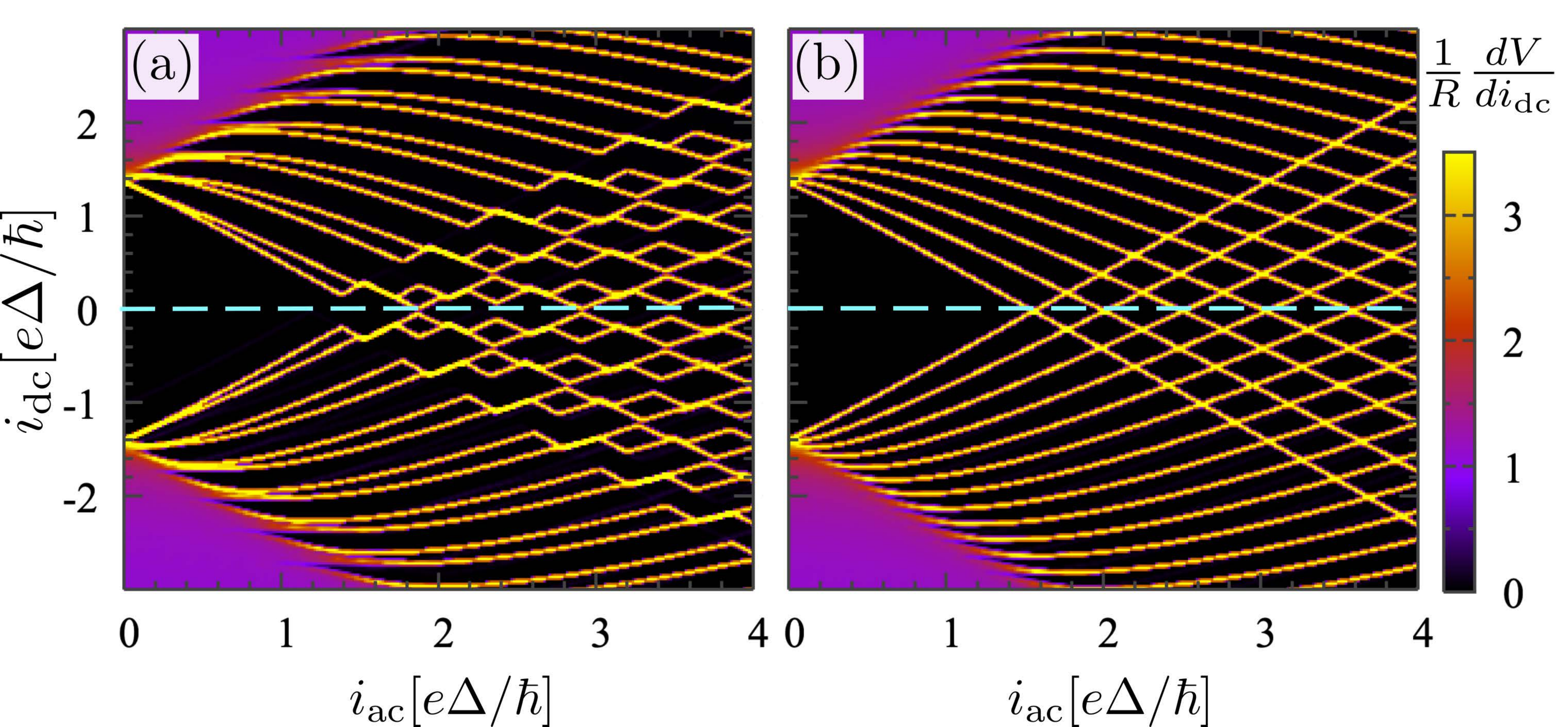}
\caption{Differential resistance, $\text{d}V/\text{d}i_{\text{dc}}$, versus $i_{\rm ac}$ and $i_{\rm dc}$ for $\Phi/\Phi_{0}=1/2$ and $(\tau_{1},\tau_{2})=(0.99,0.96)$ in panel (a) and $(\tau_{1},\tau_{2})=(0.99,0.99)$ in panel (b), showing the appearance
of additional, fractional Shapiro steps. 
}\label{S7}
\end{figure}

\section{Fractional Shapiro steps}
In the main text, we argued that in a supercurrent 
interferometer comprised of QPC JJs that contain higher harmonic content in their CPR, the first harmonic can be cancelled when
the transmissions of the two JJs are balanced by local gates ($\tau_{1}=\tau_{2}$) and a half-flux quantum ($\Phi=\Phi_{0}/2$) pierces through the interferometer loop.
The dominant contribution to the supercurrent is then 
due to the coherent tunneling of pairs of Cooper-pairs. 

To reveal this unusual tunneling effect of pairs of Cooper-pairs, we suggested in the main text the use of an AC drive that should give rise to a doubling of the Shapiro steps. 
While we showed this doubling of the Shapiro steps with the
help of the $i_{dc}-V$ curve in the main text, it can also be revealed by plotting the differential resistance, $\text{d}V/\text{d}i_{\text{dc}}$, versus $i_{\rm ac}$ and $i_{\rm dc}$. Here, we provide this additional information in Fig.\,\ref{S7}. When approaching the  balanced situation, the the differential resistance peaks split. In the balanced configuration, ($\tau_1=\tau_2$ and $\Phi=\Phi_{0}/2$) the peaks have doubled, showing the same height.

Finally, we would like to point out that is is possible to engineer junctions where the leading contribution to the supercurrent arises from the tunneling of $n>2$ Cooper pairs. The criterion is to use at least $n$ JJs with tunable transmission and magnetic flux. The interferometer CPR in this case is given by
\begin{equation}
    I(\varphi)=\sum_{m=0}^{n-1}I_m(\varphi-2m\pi\Phi/n\Phi_0)\,,
\end{equation}
where $I$ is given by Eq. (4) of the main text and the transmission of every CPR has to be equal.

\end{document}